\documentclass[journal]{IEEEtran}

\usepackage{amsmath,amssymb,amsthm}
\usepackage{bm}
\usepackage{booktabs}
\usepackage{cite}
\usepackage{graphicx}
\usepackage[hidelinks]{hyperref}

\graphicspath{{figures/}}

\theoremstyle{definition}

\begin{document}

\title{An Energy-Based Framework for Transient Stability of Grid-Forming Converter Networks With Current Limiting}

\author{Wei~Zhou,~\IEEEmembership{Student Member,~IEEE,}
and Yunjie~Gu,~\IEEEmembership{Senior Member,~IEEE}%
\thanks{This research is partly sponsored by Royal Society under award
ICA\textbackslash{}R2\textbackslash{}242103.}%
\thanks{Wei Zhou and Yunjie Gu are with Imperial College London, London, U.K.
(e-mail: w.zhou25@imperial.ac.uk; yunjie.gu@imperial.ac.uk).}}

\maketitle

\begin{abstract}
Existing analyses for the transient stability of grid-forming (GFM) converters under current
limiting mainly address single-converter systems, while network-level
transient stability analysis remains challenging. This paper develops a
network energy construction that incorporates current limiting while
retaining a provable dissipation property. This structure enables
Lyapunov-based analysis consistent with classical direct method for synchronous-machine based systems, without switching between mode-dependent energy functions.
A two-GFM example using a reactive circular current limiter and
impedance insertion illustrates the proposed framework.
Electromagnetic transient simulations
support the theoretical results.
\end{abstract}


\section*{Nomenclature}

Scalar symbols are italic; vector and matrix symbols are bold italic.
An overbar denotes a complex phasor, and bold overbarred symbols denote
vectors of phasors. Superscripts $*$ and $\mathsf T$ denote complex
conjugation and transpose, respectively.
Phasors and internal angles are expressed in a synchronous reference frame
rotating at the equilibrium angular frequency $\omega_0$.

\begin{IEEEdescription}[\IEEEusemathlabelsep\IEEEsetlabelwidth{$\Delta\bar V_k,\,\Delta V_k$}]
\item[$N$] Number of GFM buses.
\item[$i,j,k$] Bus indices.
\item[$\bar E_k,\bar V_k,\bar I_k$] Phasors of GFM $k$: internal voltage, terminal voltage, and terminal current.
\item[$E_k,V_k,I_k$] Magnitudes of $\bar E_k$, $\bar V_k$, and $\bar I_k$.
\item[$\bar{\bm E},\bar{\bm V},\bar{\bm I}$] Network vectors of the corresponding phasors.
\item[$\bar I_{kj}$] Line-current phasor directed from bus $k$ to bus $j$.
\item[$\Delta\bar V_k$] Port voltage-drop phasor, $\bar E_k-\bar V_k$.
\item[$\Delta V_k$] Port voltage-drop magnitude, $|\Delta\bar V_k|$.
\item[$\phi_k$] Phase lag of the port current relative to the port voltage drop.
\item[$f_k$] Port current-magnitude limiting function,\\
$I_k=f_k(\Delta V_k)$.
\item[$I_{\max,k}$] Current limit of GFM $k$.
\item[$X_{0,k},X_{\mathrm{eq},k}$] Baseline and equivalent inductive port reactances.
\item[$X_{ij}$] Reactance of the line connecting buses $i$ and $j$.
\item[$\theta_k$] Internal angle of GFM $k$ in the synchronous reference frame.
\item[$\bm\theta$] Network vector of internal angles.
\item[$\bm\theta_0$] Equilibrium internal-angle vector.
\item[$\omega_0$] Equilibrium angular frequency.
\item[$\omega_k$] Angular-frequency deviation of GFM $k$ from $\omega_0$.
\item[$\bm\omega$] Network vector of angular-frequency deviations from synchronism.
\item[$H_k,M_k,D_k$] Equivalent inertia constant, effective inertia coefficient, and damping coefficient of GFM $k$.
\item[$\bm M,\bm D$] Diagonal matrices of $M_k$ and $D_k$.
\item[$P_{\mathrm e,k},\bm P_{\mathrm e}$] Terminal active-power injection and its network vector.
\item[$\bm P^\star,\bm P_0$] Active-power set points and effective inputs in the synchronous frame.
\item[$U_k$] Port potential, $\int_0^{\Delta V_k}f_k(s)\,\mathrm ds$.
\item[$U_{\mathrm{net}}$] Network potential, including port and line contributions.
\item[$U_{\mathrm{eff}}$] Effective network potential obtained by eliminating terminal voltages through Kirchhoff's current law.
\item[$U_{\mathrm{tot}}$] Total potential energy.
\item[$\mathcal E$] Total energy function.
\end{IEEEdescription}

\section{Introduction}

Current limiting protects grid-forming (GFM) converters during large
disturbances but also alters their power transfer and synchronization
behavior. Transient stability analysis must therefore cover the entire
disturbance and recovery process, including normal operation,
current-limit activation, current-limited operation, and the transition
back to normal operation.

When current limiting becomes active, the converter's electrical
behavior changes, and the conventional power--angle model for unsaturated
operation no longer applies~\cite{huang2019current}.
To account for this change, previous studies have derived power--angle
models under current
limiting~\cite{rokrok2022saturation,fan2022circular,zhang2023currentconstrained,lyu2025critical},
developed equal-area-like criteria and region-of-attraction
estimates~\cite{zhang2023currentconstrained}, and examined the conditions
for post-fault recovery to normal
operation~\cite{fan2023recovery,lyu2024transition,arjomandi2025recovery}.
These analyses primarily consider a single GFM converter connected to an
infinite bus. Extending them to interconnected GFM networks is not
straightforward, because terminal voltages and power injections are jointly
determined by the network, and individual power--angle relations do not
directly yield a common network energy function.

At the network level, existing analyses impose a restrictive condition
that system trajectories avoid entering or re-entering current saturation,
enforced through energy
thresholds~\cite{li2025controllerlimits,wang2025multiparallel}.
Consequently, their energy functions characterize operation outside the
current-limited region, leaving the network energy structure during
current saturation unaddressed.
Current limiting is included in the formulation
in~\cite{baeckeland2025largesignal} via an equivalent-circuit construction
that combines the energy contributions of control and physical subsystems,
but the resulting composite energy function is not guaranteed to be
dissipative during current-limited operation. Its validity analysis relies
on approximations near an unsaturated equilibrium.
Thus, existing network-level energy methods involve either restrictions
on their applicability or compromises in their dissipation guarantees
under current limiting.

This paper develops a network energy construction that incorporates
current limiting while retaining a provable dissipation property.
Current limiting is represented by a single current--voltage characteristic
at each admissible GFM port, from which a port potential is obtained by
integration. These port potentials provide the basis for a common network
energy function with a nonpositive time derivative within the valid
reduced-model domain. Physically, current limiting reshapes the
potential governing network power transfer, while synchronization damping
dissipates transient energy. The same energy function therefore supports
stability analysis throughout normal operation, limiter activation,
current-limited operation, and recovery, without switching between
mode-dependent energy functions.

The paper is organised as follows.
Section~\ref{sec:network_potential} develops the network potential and
transient energy function for current-limited GFMs.
Section~\ref{sec:energy_stability} applies the construction to a two-GFM
system and analyzes its region of attraction and stability margins.
Section~\ref{sec:simulation_validation} presents electromagnetic transient
(EMT) validation, and
Section~\ref{sec:conclusion} concludes the paper.

\begin{figure}[t]
  \centering
  \includegraphics[width=\columnwidth]{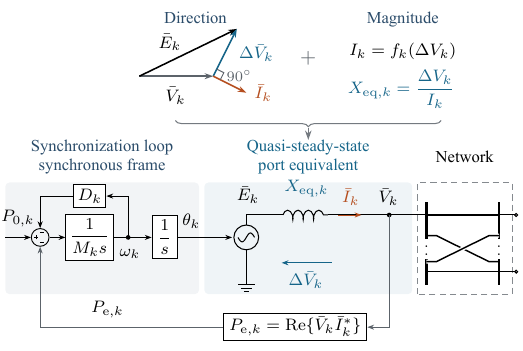}
  \caption{GFM synchronization loop in the equilibrium reference frame and
  quasi-steady-state port equivalent.}
  \label{fig:gfm_control_port}
\end{figure}

\section{Energy Structure of Current-Limited GFM Networks}
\label{sec:network_potential}

In this section, we show how preserving an inductive GFM port characteristic
before and during current limiting enables a common construction of a
network potential. After eliminating the algebraic bus voltages through
Kirchhoff's current law (KCL), the reduced network potential has an angle
gradient equal to the
vector of electrical active powers. This gradient structure enables the
construction of a transient energy function with a provable dissipation
property, connecting the proposed formulation to classical synchronous-machine
energy methods.

\subsection{GFM Port Characteristics}
\label{subsec:radial_limiter}

We focus on transient angle stability and neglect electromagnetic
transients, modelling the network and GFM ports through algebraic
phasor relations.

Consider a connected inductive network with GFMs at some buses and zero
current injections at the remaining buses. Kron reduction eliminates
the zero-injection buses while preserving the current--voltage relations
at the GFM terminal buses~\cite{dorfler2013kron}. The resulting network has
one GFM at each bus and equivalent inductive lines $(i,j)$ of
reactance $X_{ij}>0$. The following formulation refers to this reduced
network. The internal voltage is
$\bar E_k=E_ke^{\mathrm j\theta_k}$, where $E_k>0$ is held constant and
$\theta_k$ is generated by the active-power controller.
Let $\Delta\bar V_k=\bar E_k-\bar V_k$ and $\Delta V_k=|\Delta\bar V_k|$.
The port current-limiting characteristics are modelled as
\begin{equation}
    \bar I_k=e^{-\mathrm j\phi_k}f_k(\Delta V_k)
             \frac{\Delta\bar V_k}{\Delta V_k},
    \qquad \Delta V_k>0 .
    \label{eq:radial_limiter}
\end{equation}
Here, $\phi_k$ determines the phase lag of the current relative to the
port voltage drop, while the scalar function $f_k$ sets its magnitude,
$I_k=|\bar I_k|=f_k(\Delta V_k)$. This separates the phase relation from
the magnitude-limiting characteristic.

The port current-magnitude limiting function $f_k$ is continuous and
nondecreasing, with $f_k(0)=0$ and $f_k(\Delta V_k)>0$ for $\Delta V_k>0$;
the current is zero at $\Delta V_k=0$.

For the conventional circular current limiter analysed
in~\cite{fan2022circular}, the saturated inner control loop is equivalent
to a resistor, whose phase relation corresponds to $\phi_k=0$.
Grid-code guidance for GFMs describes a predominantly inductive
voltage-source response and specifies that current limiting preserve the
angle of the corresponding unlimited current phasor~\cite{entsoe2025gridforming}.
Consistent with this description, the following analysis adopts the
purely inductive case $\phi_k=\pi/2$. The phase factor then reduces to
$e^{-\mathrm j\pi/2}=-\mathrm j$, preserving the inductive phase relation
as the current magnitude is limited.

Under this phase relation, \eqref{eq:radial_limiter} is equivalently
written as
\begin{equation}
\begin{aligned}
    \bar I_k&=\frac{\Delta\bar V_k}{\mathrm jX_{\mathrm{eq},k}},\\
    X_{\mathrm{eq},k}&=\frac{\Delta V_k}{f_k(\Delta V_k)}
                     =\frac{\Delta V_k}{I_k},\qquad I_k>0.
\end{aligned}
    \label{eq:effective_port_reactance}
\end{equation}
The port therefore represents an internal voltage source behind a
variable effective reactance. In this phasor representation,
this is the same inductive relation as impedance insertion using a
purely inductive virtual reactance. When $f_k$ is invertible, the effective reactance can
be expressed as a function of current,
$X_{\mathrm{eq},k}(I_k)=f_k^{-1}(I_k)/I_k$.

Fig.~\ref{fig:gfm_control_port} illustrates the phase and magnitude
relations together with the synchronization loop. The loop supplies the
internal angle, while the terminal active power closes the feedback loop.
Its effective power input $P_{0,k}$ accounts for the equilibrium frequency
offset, as defined in Section~\ref{subsec:multi_gfm_energy}.

For example, a reactive circular current limiter applied to the command
$-\mathrm j\Delta\bar V_k/X_{0,k}$, where $X_{0,k}>0$ is the baseline
total port reactance, gives
\begin{equation}
    f_k(\Delta V_k)=\min\left\{\frac{\Delta V_k}{X_{0,k}},\,I_{\max,k}\right\}.
    \label{eq:hard_radial_characteristic}
\end{equation}
Below $\Delta V_k=X_{0,k}I_{\max,k}$, the equivalent reactance is $X_{0,k}$.
Above this threshold, it increases as $\Delta V_k/I_{\max,k}$ to maintain the
current magnitude. On this saturated branch, the effective reactance
varies with the voltage drop at fixed current and cannot be determined
by current alone. Impedance insertion using a purely inductive
virtual reactance can instead be described by
$\Delta V_k=I_kX_{\mathrm{eq},k}(I_k)$. An increasing positive
$X_{\mathrm{eq},k}(I_k)$ gives a single-valued $I_k=f_k(\Delta V_k)$, without necessarily imposing
a strict current ceiling.

For the inductive port characteristics considered here, the phase
relation is preserved throughout normal operation, current limiting,
and recovery. The GFM ports and network lines therefore share the same
current--voltage phase structure, allowing their potentials to be
combined into a network potential, as developed in the next subsection.

\subsection{Network Potential and Active Power Gradient Structure}
\label{subsec:network_gradient}

For the inductive port relation with $\phi_k=\pi/2$, the
current-magnitude characteristic $f_k$ defines the port potential
\begin{equation}
    U_k(\Delta V_k)=\int_0^{\Delta V_k}f_k(s)\,\mathrm ds.
    \label{eq:converter_potential}
\end{equation}
Since current limiting is incorporated in $f_k$, the same potential
describes the port across current-limit activation and deactivation.
For an inductive line between buses $i$ and $j$, the current magnitude is
$\Delta V_{ij}/X_{ij}$, where $\Delta V_{ij}=|\bar V_i-\bar V_j|$.
Following the same construction as in \eqref{eq:converter_potential},
its potential is
\begin{equation}
    U_{ij}(\Delta V_{ij})
      =\int_0^{\Delta V_{ij}}\frac{s}{X_{ij}}\,\mathrm ds
      =\frac{\Delta V_{ij}^2}{2X_{ij}}.
    \label{eq:line_potential}
\end{equation}
Combining the port and line potentials gives
\begin{equation}
    U_{\mathrm{net}}(\bm\theta,\bar{\bm V})
    =\sum_k U_k\bigl(|\bar E_k-\bar V_k|\bigr)
    +\sum_{(i,j)}\frac{|\bar V_i-\bar V_j|^2}{2X_{ij}},
    \label{eq:network_potential}
\end{equation}
where the second sum includes each unordered pair of distinct buses
once. For buses with no direct connection in the reduced network,
we adopt the convention $1/X_{ij}=0$, equivalently $X_{ij}=\infty$.

For given internal angles, let $\bar{\bm V}(\bm\theta)$ denote a solution
of the algebraic network equations. Where these equations are smooth and
their terminal-voltage Jacobian is nonsingular, the implicit function
theorem provides a differentiable local voltage map.
Substituting the voltage solution into \eqref{eq:network_potential} defines the
effective network potential
\begin{equation}
    U_{\mathrm{eff}}(\bm\theta)
      =U_{\mathrm{net}}\bigl(\bm\theta,\bar{\bm V}(\bm\theta)\bigr).
    \label{eq:reduced_angle_potential}
\end{equation}
Its angle dependence arises both explicitly through the internal
voltages and implicitly through the terminal voltages. On a differentiable
voltage solution branch, applying the chain rule gives
\begin{equation}
    \frac{\partial U_{\mathrm{eff}}}{\partial\bm\theta}
      =\left.\frac{\partial U_{\mathrm{net}}}{\partial\bm\theta}
        \right|_{\bar{\bm V}}
        +\frac{\partial U_{\mathrm{net}}}{\partial\bar{\bm V}}
        \frac{\partial\bar{\bm V}}{\partial\bm\theta}
        +\frac{\partial U_{\mathrm{net}}}{\partial\bar{\bm V}^{*}}
        \frac{\partial\bar{\bm V}^{*}}{\partial\bm\theta},
    \label{eq:reduced_potential_chain_rule}
\end{equation}
where we evaluate the derivatives using Wirtinger calculus, writing
$|z|=\sqrt{zz^*}$ and treating $z$ and $z^*$ as formally independent
variables when taking partial derivatives~\cite{hjorungnes2007complex}.

Let $\bar I_{kj}=-\mathrm j(\bar V_k-\bar V_j)/X_{kj}$ denote
the line current from bus $k$ to an adjacent bus $j$.
Differentiating \eqref{eq:network_potential} with respect to the terminal
voltages yields
\begin{equation}
\begin{aligned}
    \frac{\partial U_{\mathrm{net}}}{\partial\bar V_k}
      &=-\frac{f_k(\Delta V_k)}{2\Delta V_k}
        (\bar E_k-\bar V_k)^*
        +\sum_j\frac{(\bar V_k-\bar V_j)^*}{2X_{kj}}\\
      &=\frac{\mathrm j}{2}
        \left(\bar I_k-\sum_j\bar I_{kj}\right)^*=0,\\[3pt]
    \frac{\partial U_{\mathrm{net}}}{\partial\bar V_k^*}
      &=\left(\frac{\partial U_{\mathrm{net}}}{\partial\bar V_k}\right)^*
        =0.
\end{aligned}
    \label{eq:kcl_stationarity}
\end{equation}
The first derivative vanishes by KCL, $\bar I_k=\sum_j\bar I_{kj}$,
and the conjugate derivative vanishes because $U_{\mathrm{net}}$ is
real-valued. Expressions at $\Delta V_k=0$ are interpreted by continuity.

Consequently, both terminal-voltage terms in
\eqref{eq:reduced_potential_chain_rule} vanish. With the terminal voltages
held fixed, only the $k$th port potential depends on $\theta_k$.
Using $\partial\bar E_k/\partial\theta_k=\mathrm j\bar E_k$, we obtain
\begin{equation}
\begin{aligned}
    \frac{\partial U_{\mathrm{eff}}}{\partial\theta_k}
      &=\left.\frac{\partial U_{\mathrm{net}}}{\partial\theta_k}
        \right|_{\bar{\bm V}}\\
      &=\frac{f_k(\Delta V_k)}{\Delta V_k}
        \operatorname{Re}\left\{
        (\bar E_k-\bar V_k)^*\,\mathrm j\bar E_k\right\}\\
      &=\operatorname{Re}\{\bar E_k\bar I_k^*\}=P_{\mathrm e,k},\\[3pt]
    \frac{\partial U_{\mathrm{eff}}}{\partial\bm\theta}
      &=\bm P_{\mathrm e}^{\mathsf T}.
\end{aligned}
    \label{eq:power_gradient}
\end{equation}
The inductive port absorbs no active power, so
$\operatorname{Re}\{\bar E_k\bar I_k^*\}
=\operatorname{Re}\{\bar V_k\bar I_k^*\}$ is the terminal active-power
injection.

Continuity of $f_k$ makes the port potential continuously differentiable,
including at hard-limiter knees. If the terminal-voltage Jacobian becomes
singular, the terminal voltages may be nonunique. For the nondecreasing
port characteristics considered here, it can be shown that all KCL
solutions at given internal angles yield the same network potential and
active powers, and that the effective potential remains continuously
differentiable with the gradient in \eqref{eq:power_gradient}.
Such voltage nonuniqueness nevertheless indicates an unregulated voltage
degree of freedom in the algebraic model. If all GFMs are current-limited
and such a degree of freedom remains, an additional voltage-regulation
mechanism is required, since transient angle stability alone does not
ensure voltage stability.

Under current limiting, a closed-form expression for
$U_{\mathrm{eff}}(\bm\theta)$ is generally unavailable, but it can be
evaluated numerically. The gradient relation in \eqref{eq:power_gradient}
allows potential differences to be computed by integrating
$\bm P_{\mathrm e}$ along a path in angle space. A more direct approach
is to solve the KCL equations for the terminal voltages at the given
$\bm\theta$ and substitute the solution into
\eqref{eq:network_potential}, as prescribed by
\eqref{eq:reduced_angle_potential}. This evaluates the effective potential
by summing the port and line potentials; the port potentials in
\eqref{eq:converter_potential} can be computed in advance analytically
or by numerical integration.

The active-power gradient structure in \eqref{eq:power_gradient} enables
the construction of a network transient energy function with a provable dissipation property,
as developed in the following subsection.

\subsection{Transient Energy Function}
\label{subsec:multi_gfm_energy}

The active-power gradient relation in \eqref{eq:power_gradient} enables
the network potential to be combined with the kinetic energy of the
synchronization dynamics, yielding a transient energy function with a
provable dissipation property. In the synchronous reference frame,
these dynamics are
\begin{equation}
    \dot{\bm\theta}=\bm\omega,\qquad
    \bm M\dot{\bm\omega}
    =\bm P_0-\bm P_{\mathrm e}(\bm\theta)-\bm D\bm\omega,
    \label{eq:multi_gfm_dynamics}
\end{equation}
where $\bm\omega$ contains angular-frequency deviations from $\omega_0$
in rad/s. The positive diagonal matrices $\bm M=\operatorname{diag}(M_k)$
and $\bm D=\operatorname{diag}(D_k)$ contain the effective inertia and
damping coefficients on a common per-unit power base.
A uniform angle rotation changes neither voltage-drop magnitudes nor
power transfer. Hence, $U_{\mathrm{eff}}$ depends only on relative angles,
and $\bm1^{\mathsf T}\bm P_{\mathrm e}=0$ for this lossless network.

For controllers sharing the same nominal frequency reference, the
equilibrium frequency offset from that reference is
$(\bm1^{\mathsf T}\bm P^\star)/(\bm1^{\mathsf T}\bm D\bm1)$.
Absorbing the corresponding damping offset into the power inputs gives
\begin{equation}
    \bm P_0=\bm P^\star-\bm D\bm1\,
    \frac{\bm1^{\mathsf T}\bm P^\star}{\bm1^{\mathsf T}\bm D\bm1},
    \qquad \bm1^{\mathsf T}\bm P_0=0.
    \label{eq:effective_power_inputs}
\end{equation}
Thus, Fig.~\ref{fig:gfm_control_port} uses $P_{0,k}$ as its input when
$\omega_k$ is measured from $\omega_0$. The electrical power
$P_{\mathrm e,k}$ is unchanged by this reference-frame transformation.
For balanced set points, $\bm P_0=\bm P^\star$ and $\omega_0$ equals the
nominal reference frequency.

Let $\bm\theta_0$ denote an equilibrium angle vector satisfying
$\bm P_{\mathrm e}(\bm\theta_0)=\bm P_0$.
Following the kinetic-plus-angle-potential form of transient energy
methods~\cite{shuai2019transient}, define
\begin{equation}
\begin{aligned}
    \mathcal E
    &=\tfrac12\bm\omega^{\mathsf T}\bm M\bm\omega
       +U_{\mathrm{tot}}(\bm\theta),\\
    U_{\mathrm{tot}}(\bm\theta)
    &=U_{\mathrm{eff}}(\bm\theta)-U_{\mathrm{eff}}(\bm\theta_0)
      -\bm P_0^{\mathsf T}(\bm\theta-\bm\theta_0).
\end{aligned}
    \label{eq:multi_gfm_energy}
\end{equation}
The first term measures the frequency excursion weighted by effective
inertia. The second accounts for the electrical power relative to the
effective power inputs: $\nabla U_{\mathrm{tot}}=\bm P_{\mathrm e}-\bm P_0$.

Using \eqref{eq:power_gradient},
$\dot U_{\mathrm{tot}}=(\bm P_{\mathrm e}-\bm P_0)^{\mathsf T}\bm\omega$.
Substitution of \eqref{eq:multi_gfm_dynamics} into the derivative of
\eqref{eq:multi_gfm_energy} cancels the power-transfer terms, giving
\begin{equation}
    \dot{\mathcal E}=-\bm\omega^{\mathsf T}\bm D\bm\omega\leq0 .
    \label{eq:multi_gfm_energy_derivative}
\end{equation}
Thus, the network energy structure remains dissipative with current
limiting incorporated.

For stability assessment, take $(\bm\theta_0,\bm0)$ to be a target
stable equilibrium point (SEP), at which $U_{\mathrm{tot}}$ has a strict
local minimum after removing the uniform-angle degree of freedom.
Suppose the potential barriers bounding its well are attained at
unstable equilibrium points (UEPs), and let $\mathcal U_{\mathrm b}$
denote their angle vectors. The lowest barrier defines the critical energy
\begin{equation}
    \mathcal E_{\mathrm{cr}}
      =\min_{\bm\theta_{\mathrm u}\in\mathcal U_{\mathrm b}}
       U_{\mathrm{tot}}(\bm\theta_{\mathrm u}).
    \label{eq:network_critical_energy}
\end{equation}
Consider the energy sublevel set
\begin{equation}
    \mathcal E(\bm\theta,\bm\omega)<\mathcal E_{\mathrm{cr}}.
    \label{eq:network_energy_criterion}
\end{equation}
Its connected component containing the target SEP is positively
invariant by \eqref{eq:multi_gfm_energy_derivative}. If this component
is bounded in relative-angle coordinates, its closure lies within the
model domain, and it contains no other equilibrium, positive damping
and LaSalle's invariance principle guarantee convergence to the target
SEP. This component therefore provides a conservative estimate of its
region of attraction (ROA). Section~\ref{sec:energy_stability}
illustrates this network criterion using a two-GFM example.

\begin{figure}[!t]
\centering
  \includegraphics[width=\columnwidth]{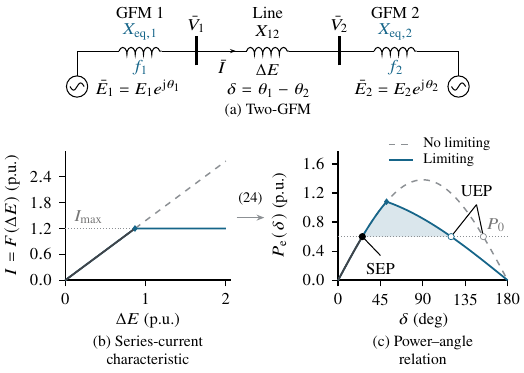}
  \caption{Two-GFM network potential and power--angle interpretation.
  (a) Equivalent circuit. (b) Series-current characteristics $I=F(\Delta E)$.
  (c) Power--angle curves obtained from Fig.~\ref{fig:two_gfm_structure}(b) using
  \eqref{eq:two_gfm_power_current}, whose areas
  relative to $P_0$ represent changes in total potential energy.
  The stable equilibrium point (SEP) and the boundary unstable equilibrium
  point (UEP) are marked in Fig.~\ref{fig:two_gfm_structure}(c).
  The curves labelled ``Limiting'' use the reactive circular current limiter
  in \eqref{eq:hard_radial_characteristic} at both GFMs.}
  \label{fig:two_gfm_structure}
\vspace{12pt}
  \includegraphics[width=\columnwidth]{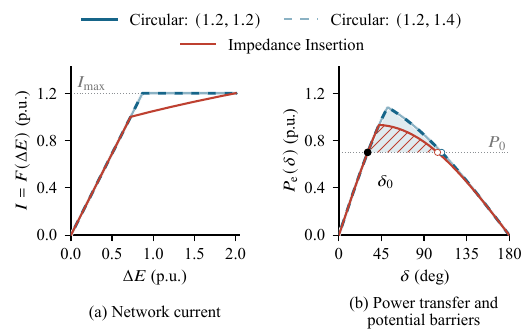}
  \caption{Current utilization and potential barriers with a reactive circular
  current limiter and impedance insertion.
  (a) Series-current characteristics $I=F(\Delta E)$.
  (b) Power--angle curves obtained using \eqref{eq:two_gfm_power_current}.
  The circular current limiter legend entries specify $(I_{\max,1},I_{\max,2})$
  in p.u.; their curves coincide. Blue shading and red hatching show the
  potential barriers for the circular current limiter and impedance insertion,
  respectively. The filled circle marks the common SEP; open circles mark the
  boundary UEPs. The network and impedance insertion parameters are as in
  Section~\ref{sec:simulation_validation}, with $P_0=0.7$ p.u.
  Both limiters reach $I=1.2$ p.u. at $\Delta E=2$ p.u.}
  \label{fig:limiter_comparison}
\end{figure}
\section{Transient Stability Assessment via a Two-GFM Example}
\label{sec:energy_stability}

Building on the dissipative energy structure and stability criterion in
Section~\ref{sec:network_potential}, this section assesses transient
stability using the two-GFM network in Fig.~\ref{fig:two_gfm_structure}(a).
Even in this simple setting, the energy structure reveals how
current-limited ports jointly shape power transfer and how current
utilization and dissipation affect transient stability.

\subsection{Network Potential and Power Transfer}
\label{subsec:two_gfm_electrical}

For the two-GFM network in Fig.~\ref{fig:two_gfm_structure}(a), the
effective network potential in \eqref{eq:reduced_angle_potential} becomes
\begin{equation}
    U_{\mathrm{eff}}(\theta_1,\theta_2)
      =U_1(\Delta V_1)+U_2(\Delta V_2)+\tfrac12X_{12}I^2,
    \label{eq:two_gfm_reduced_potential}
\end{equation}
where the port voltage drops and common current magnitude are evaluated
at the network solution for $(\theta_1,\theta_2)$.
A common angle rotation leaves this potential unchanged, so it depends
only on the relative angle $\delta=\theta_1-\theta_2$ and can be written
as $U_{\mathrm{eff}}(\delta)$. Applying
\eqref{eq:power_gradient} gives
\begin{equation}
    P_{\mathrm e}(\delta)=P_{\mathrm e,1}
      =\frac{\partial U_{\mathrm{eff}}}{\partial\theta_1}
      =\frac{\mathrm dU_{\mathrm{eff}}(\delta)}{\mathrm d\delta}.
    \label{eq:relative_power_potential}
\end{equation}
For the two-GFM system, the effective potential can therefore be
evaluated by integrating the power--angle characteristic. Taking the
target SEP at $\delta_0$ as the reference gives
\begin{equation}
    U_{\mathrm{eff}}(\delta)-U_{\mathrm{eff}}(\delta_0)
      =\int_{\delta_0}^{\delta}P_{\mathrm e}(\sigma)\,\mathrm d\sigma.
    \label{eq:two_gfm_power_integral}
\end{equation}

With $P_0=P_{0,1}=-P_{0,2}$ as defined in
Section~\ref{subsec:multi_gfm_energy} and
$P_{\mathrm e}(\delta_0)=P_0$, the total potential in
\eqref{eq:multi_gfm_energy} follows from
\eqref{eq:two_gfm_power_integral} as
\begin{equation}
\begin{aligned}
    U_{\mathrm{tot}}(\delta)
      &=U_{\mathrm{eff}}(\delta)-U_{\mathrm{eff}}(\delta_0)
        -P_0(\delta-\delta_0)\\
      &=\int_{\delta_0}^{\delta}
        [P_{\mathrm e}(\sigma)-P_0]\,\mathrm d\sigma.
\end{aligned}
    \label{eq:shifted_angle_potential}
\end{equation}
The signed area between $P_{\mathrm e}(\delta)$ and $P_0$ thus measures
the total potential relative to the equilibrium. For increasing $\delta$,
accelerating regions with $P_0>P_{\mathrm e}$ lower this potential,
whereas decelerating regions with $P_{\mathrm e}>P_0$ raise it.
In the absence of damping, the exchange between potential and kinetic
energy gives the classical equal-area criterion.
The dissipation relation \eqref{eq:multi_gfm_energy_derivative} adds
the effect of damping to this energy interpretation, providing the
basis for the stability assessment below.

To determine this power--angle characteristic, let
$\Delta E=|\bar E_1-\bar E_2|$ denote the internal-voltage
separation. The common current magnitude $I$ and the port voltage
drops satisfy the network equations
\begin{equation}
\begin{cases}
    f_1(\Delta V_1)=f_2(\Delta V_2)=I,\\
    \Delta V_1+X_{12}I+\Delta V_2=\Delta E.
\end{cases}
    \label{eq:two_gfm_series_current}
\end{equation}
Since $f_1$ and $f_2$ are continuous and nondecreasing and
$X_{12}>0$, \eqref{eq:two_gfm_series_current} uniquely determines
the common current for each $\Delta E$. We write this as
\begin{equation}
    I=F(\Delta E).
    \label{eq:network_current_characteristic}
\end{equation}
The angle dependence enters through
\begin{equation}
    \Delta E(\delta)=\sqrt{E_1^2+E_2^2-2E_1E_2\cos\delta}.
    \label{eq:two_gfm_voltage_separation}
\end{equation}
The inductive series phase relation then gives
\begin{equation}
    P_{\mathrm e}(\delta)
      =\frac{E_1E_2\sin\delta}{\Delta E(\delta)}
       F\bigl(\Delta E(\delta)\bigr),
    \label{eq:two_gfm_power_current}
\end{equation}
where the value at $\Delta E=0$ is defined by taking the limit as
$\Delta E\to0$.

Without current limiting, the port reactances are fixed at $X_{0,k}$.
With $X_\Sigma=X_{0,1}+X_{12}+X_{0,2}$,
\eqref{eq:two_gfm_series_current} has an explicit solution, yielding
\begin{equation}
    F(\Delta E)=\frac{\Delta E}{X_\Sigma},\qquad
    P_{\mathrm e}(\delta)=\frac{E_1E_2}{X_\Sigma}\sin\delta.
    \label{eq:classical_two_gfm_power}
\end{equation}
Thus, the familiar sinusoidal power--angle relation is recovered as the
unlimited case. With current limiting, the current characteristic
$F$ changes the power--angle curve while preserving its
relation to the network potential
in \eqref{eq:relative_power_potential}.
Fig.~\ref{fig:two_gfm_structure}(c) illustrates how current limiting
reshapes the power--angle curve and the associated potential barrier.

\subsection{Energy-Based Transient Stability Assessment}
\label{subsec:roa_fault_clearing}

For the two-GFM system, the total energy in
\eqref{eq:multi_gfm_energy} is
\begin{equation}
    \mathcal E
      =\tfrac12M_1\omega_1^2+\tfrac12M_2\omega_2^2
       +U_{\mathrm{tot}}(\delta).
    \label{eq:two_gfm_total_energy}
\end{equation}
Define $\omega_{\mathrm c}=(M_1\omega_1+M_2\omega_2)/(M_1+M_2)$
as the center-of-inertia (CoI) frequency deviation,
$\omega_{\delta}=\omega_1-\omega_2$ as the relative frequency, and
$M_{\delta}=M_1M_2/(M_1+M_2)$ as the effective inertia. The kinetic
energy separates into CoI and relative-motion components, so
\begin{equation}
    \mathcal E
      =\tfrac12(M_1+M_2)\omega_{\mathrm c}^2
        +\tfrac12M_{\delta}\omega_{\delta}^2
        +U_{\mathrm{tot}}(\delta).
    \label{eq:two_gfm_energy_decomposition}
\end{equation}

Assume that the two GFMs have equal damping-to-inertia ratios, and let
$\gamma=D_1/M_1=D_2/M_2>0$ and $D_{\delta}=\gamma M_{\delta}$.
Using $P_{0,1}+P_{0,2}=P_{\mathrm e,1}+P_{\mathrm e,2}=0$,
the dynamics in \eqref{eq:multi_gfm_dynamics} separate as
\begin{equation}
\begin{aligned}
    \dot\omega_{\mathrm c}&=-\gamma\omega_{\mathrm c},
      \qquad \dot\delta=\omega_{\delta},\\
    M_{\delta}\dot\omega_{\delta}
      &=P_0-P_{\mathrm e}(\delta)-D_{\delta}\omega_{\delta}.
\end{aligned}
    \label{eq:matched_relative_dynamics}
\end{equation}
The CoI motion decays independently and does not feed
the relative motion. Transient stability can therefore be assessed
using only $(\delta,\omega_{\delta})$, even when
$\omega_{\mathrm c}(0)\ne0$. Their energy is the sum of the remaining
kinetic energy and potential in
\eqref{eq:two_gfm_energy_decomposition}, with dissipation given by
\eqref{eq:multi_gfm_energy_derivative} and
\eqref{eq:matched_relative_dynamics}:
\begin{equation}
\begin{aligned}
    \mathcal E_{\delta}
      &=\tfrac12M_{\delta}\omega_{\delta}^2+U_{\mathrm{tot}}(\delta),\\
    \dot{\mathcal E}_{\delta}&=-D_{\delta}\omega_{\delta}^2.
\end{aligned}
    \label{eq:relative_energy_dissipation}
\end{equation}

The network critical energy in \eqref{eq:network_critical_energy}
can now be evaluated from the potential along $\delta$.
The power--angle curves considered here have one UEP per $2\pi$ period.
For $P_0>0$, the UEP immediately to the right of the target SEP gives
the lower barrier; denote its angle by $\delta_{\mathrm u}$.
In Fig.~\ref{fig:two_gfm_structure}(c), this barrier is the area between
$P_{\mathrm e}(\delta)$ and $P_0$ from $\delta_0$ to $\delta_{\mathrm u}$.
Using \eqref{eq:shifted_angle_potential}, the critical energy is
\begin{equation}
    \mathcal E_{\mathrm{cr}}
      =\int_{\delta_0}^{\delta_{\mathrm u}}
       \bigl[P_{\mathrm e}(\delta)-P_0\bigr]\,\mathrm d\delta.
    \label{eq:two_gfm_critical_energy}
\end{equation}
Under the conditions in Section~\ref{subsec:multi_gfm_energy}, applying
\eqref{eq:network_energy_criterion} to the relative-motion energy gives
the ROA estimate as the connected component containing $(\delta_0,0)$ of
\begin{equation}
    \mathcal E_{\delta}(\delta,\omega_{\delta})
      <\mathcal E_{\mathrm{cr}}.
    \label{eq:two_gfm_energy_criterion}
\end{equation}

\subsection{Physical Insights into Transient Stability}
\label{subsec:physical_insights}

\emph{Reactive circular current limiter:}
When both GFMs use the reactive circular current limiter in
\eqref{eq:hard_radial_characteristic}, the series relation
\eqref{eq:two_gfm_series_current} gives
\begin{equation}
    F(\Delta E)=\min\{\Delta E/X_\Sigma,I_{\max,1},I_{\max,2}\}.
    \label{eq:two_gfm_cl_current}
\end{equation}
With the baseline reactances fixed, the smaller of the two current
limits determines the saturation of $F$. Increasing only the larger
limit changes neither $F$ nor the power--angle curve in
\eqref{eq:two_gfm_power_current}. Consequently, the potential barrier,
energy-based ROA estimate, and ROA of the angle dynamics remain
unchanged when the other system parameters are fixed.
Fig.~\ref{fig:limiter_comparison} illustrates this result: the equal-
and unequal-limit cases with the circular current limiter have coincident
current and power curves.

\emph{Simultaneous current limiting:}
For equal current limits, both GFMs can enter current limiting during
the same transient. The reduction of power transfer and potential
barrier illustrated in Fig.~\ref{fig:two_gfm_structure}(c) does not
necessarily eliminate the target SEP or its ROA. At the ascending
intersection of $P_{\mathrm e}(\delta)$ with $P_0$, the potential
$U_{\mathrm{tot}}$ has a strict local minimum. Thus,
$\mathcal E_{\delta}$ is locally positive definite and nonincreasing,
establishing Lyapunov stability of the SEP. The sublevel component in
\eqref{eq:two_gfm_energy_criterion} can extend into the region where
both GFMs are current-limited, as it does for the equal-limit case with the circular current limiter
in Fig.~\ref{fig:limiter_comparison}. States in this part of the
estimate still return to the SEP under the conditions in
Section~\ref{subsec:multi_gfm_energy}, because the same dissipative
energy function applies throughout limiter activation and deactivation.

In this equal-limit case, simultaneous saturation makes the
terminal-voltage Jacobian singular: \eqref{eq:two_gfm_series_current}
fixes the sum of the port voltage drops, but not their allocation.
Neither port regulates this remaining voltage degree of freedom.
Although the relative-angle trajectory can return to the SEP, the two
terminal voltages may drift during the transient when both GFMs are
current-limited and no additional voltage-regulation mechanism is present.

\emph{Current utilization with impedance insertion:}
Impedance insertion can reduce current before it reaches the
converter's current capacity. To examine this effect, let both ports
use $X_{\mathrm{eq},k}(I)=X_{0,k}
+k_{\mathrm{ins}}\max\{I-I_{\mathrm{th}},0\}$, where
$k_{\mathrm{ins}}>0$ is the insertion gain and $I_{\mathrm{th}}$ is the activation threshold.
Substituting the port voltage drops
$\Delta V_k=f_k^{-1}(I)=IX_{\mathrm{eq},k}(I)$ into
\eqref{eq:two_gfm_series_current} gives the equation defining the series
current $I=F(\Delta E)$:
\begin{equation}
    \Delta E=I\bigl[X_\Sigma+2k_{\mathrm{ins}}
                         \max\{I-I_{\mathrm{th}},0\}\bigr].
    \label{eq:two_gfm_vi_current}
\end{equation}
Fig.~\ref{fig:limiter_comparison} compares the reactive circular current
limiter and impedance insertion at the same baseline reactances and maximum current over
the admissible voltage range.
Fig.~\ref{fig:limiter_comparison}(a) shows that impedance insertion supplies
less current at intermediate voltage
separations, leaving some current capacity unused. Through
\eqref{eq:two_gfm_power_current}, this gives the lower power--angle
curve in Fig.~\ref{fig:limiter_comparison}(b).
The boundary UEP moves closer to the common SEP,
reducing the potential barrier. For this comparison, the corresponding
energy-based ROA estimate is therefore smaller. Current utilization
throughout the transient, as well as the maximum current, determines
the stability margin.

However, with the impedance insertion characteristic considered here,
the port current continues to increase with the voltage drop after
activation, retaining a degree of voltage regulation during current
limiting. This avoids the nonunique voltage-drop allocation that can
arise when both circular current limiters saturate and can benefit
voltage stability. The comparison therefore illustrates a tradeoff:
impedance insertion offers stronger voltage regulation during current
limiting, at the cost of lower current utilization and a smaller
transient angle-stability margin.

\emph{Conservatism of the energy estimate:}
Although \eqref{eq:relative_energy_dissipation} includes damping, the
barrier-based estimate in \eqref{eq:two_gfm_energy_criterion} does not
credit the energy dissipated along the subsequent trajectory. An
initial state outside this estimate may still return to the SEP if
sufficient energy is dissipated before the trajectory crosses the
potential barrier. The estimate is therefore conservative; exclusion
from it does not establish instability.

\begin{figure*}[t]
\centering
\includegraphics[width=\textwidth]{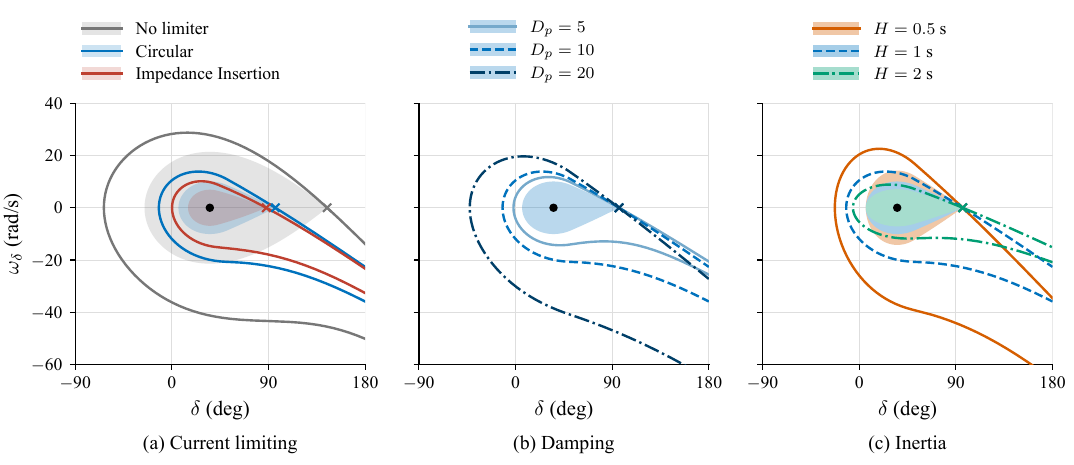}
\caption{Conservatism of the energy-based ROA estimates for the two-GFM system.
Curves show the exact ROA boundaries, computed by backward integration of
the stable manifolds of boundary UEPs; shaded regions show the corresponding
ROA estimates obtained from energy sublevel sets.
Dots mark the target SEP and crosses mark UEPs.
(a) Limiter comparison with $H=1$ s and $D_p=10$.
(b) Damping comparison with $H=1$ s; the energy estimate is the same for
all three damping values.
(c) Inertia comparison with $D_p=10$; smaller shaded estimates overlay
larger ones.
Fig.~\ref{fig:roa_energy_certificate}(b) and (c) use the reactive circular current limiter.
Other parameters are as in Section~\ref{sec:simulation_validation}.}
\label{fig:roa_energy_certificate}
\end{figure*}

Fig.~\ref{fig:roa_energy_certificate} illustrates this conservatism by
comparing the energy estimates with the exact ROAs, whose boundaries
are reconstructed from the stable manifolds of boundary UEPs.
These manifolds are obtained by
integrating \eqref{eq:matched_relative_dynamics} backward from small
perturbations along the stable eigendirections~\cite{chiang1988stability}.
The energy sublevel sets remain contained within the corresponding exact
ROAs even when current limiting is activated, as illustrated for both
the reactive circular current limiter and impedance insertion in
Fig.~\ref{fig:roa_energy_certificate}(a), verifying the validity of the
proposed framework under current limiting.
At fixed inertia and electrical parameters,
reducing damping leaves the energy estimate unchanged but brings the
exact ROA boundary closer to it, as shown in
Fig.~\ref{fig:roa_energy_certificate}(b). At fixed damping
and electrical parameters, increasing inertia changes both regions but
also narrows the gap between them, as shown in
Fig.~\ref{fig:roa_energy_certificate}(c). In these
comparisons, lower damping or larger inertia therefore makes the
energy-based estimate less conservative and more accurate, because the
impact of dissipation is lower.

\begin{figure*}[!t]
\centering
\includegraphics[width=\textwidth]{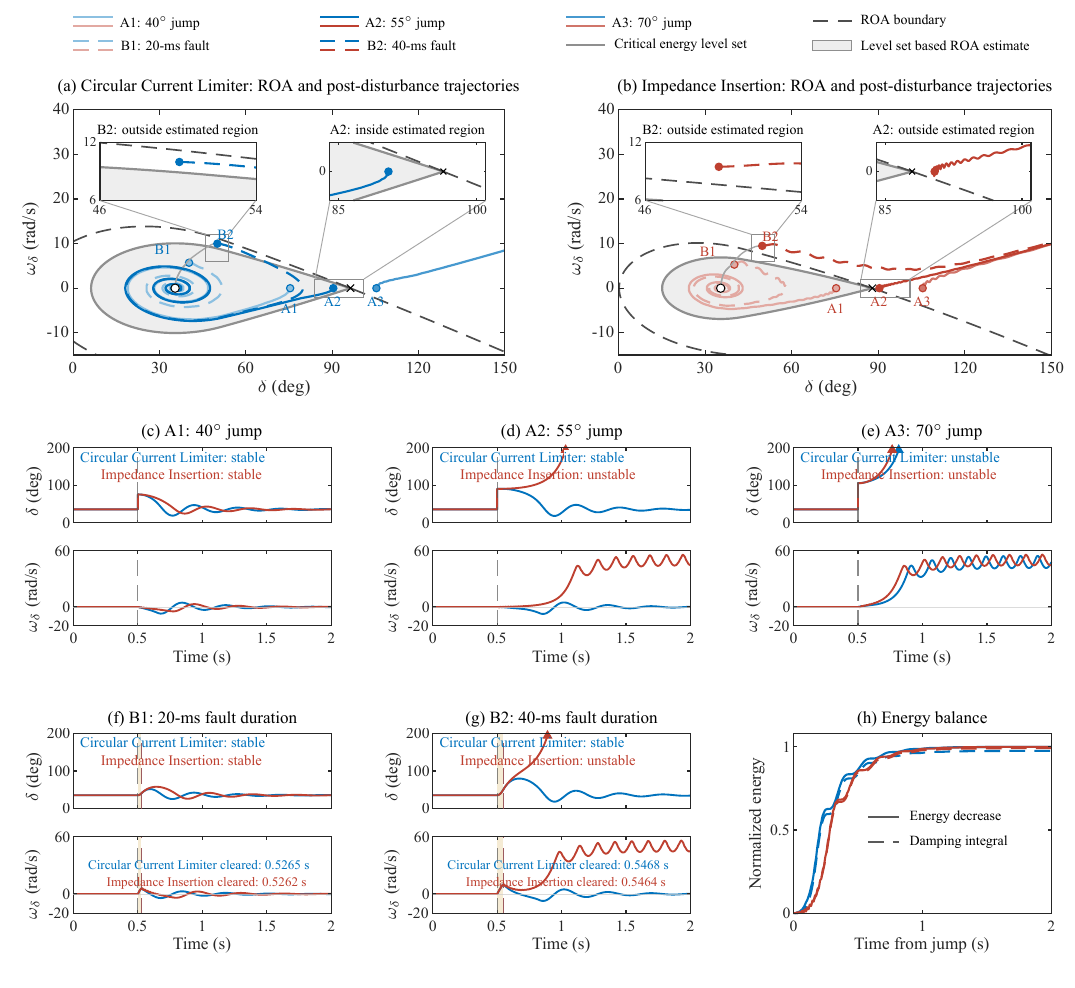}
\caption{EMT validation with a reactive circular current limiter (blue)
and impedance insertion (red).
(a), (b) ROA boundaries, energy level sets, and trajectories; gray fill
shows the inner ROA estimates. Colored circles mark post-jump or
fault-clearing states, and thin gray curves show fault-on motion.
(c)--(g) Corresponding relative-angle and frequency responses, with phase
jumps and faults applied at $0.5$ s; shading marks fault-on intervals and
triangles indicate continued angle growth.
Stable and unstable denote a return to the target SEP and loss of
synchronism, respectively.
(h) Energy decrease and damping integral for the stable $40^\circ$
jumps, normalized by post-jump energy.}
\label{fig:emt_time_domain}
\end{figure*}

\section{Simulation Validation}
\label{sec:simulation_validation}

This section uses EMT simulations in
MATLAB/Simulink to evaluate the proposed energy-based approach to
transient stability assessment of GFM converters with current limiting.
The EMT model, including the current-limiter implementations, will be
made publicly available~\cite{zhou_emt_model}.
The case studies include phase jumps and three-phase short-circuit faults.
The baseline parameters are listed in Table~\ref{tab:simulation_parameters}.
The balanced set points $P_1^\star=-P_2^\star=P_0$ give a synchronous
frequency equal to the nominal reference frequency $\omega_0$.
For each converter, $M_i=2H/\omega_0$ and $D_i=D_p/\omega_0$.
Impedance insertion uses
$X_{\mathrm{eq},k}(I)=X_{0,k}+k_{\mathrm{ins}}\max\{I-I_{\mathrm{th}},0\}$; both characteristics
reach $1.2$ p.u. at $\Delta E=2$ p.u. In the EMT model, the baseline
port reactance comprises $0.0618$ p.u. physical and $0.1$ p.u. virtual
reactance.

\begin{table}[!t]
\caption{Baseline simulation parameters}
\label{tab:simulation_parameters}
\centering\footnotesize
\setlength{\tabcolsep}{3pt}
\begin{tabular*}{\columnwidth}{@{\extracolsep{\fill}}lrlr@{}}
\specialrule{0.4pt}{0pt}{1.2pt}
\specialrule{0.4pt}{0pt}{2.5pt}
Parameter & Value & Parameter & Value\\
\midrule
$E_1,E_2$ & $1$ p.u. & $\omega_0$ & $2\pi50$ rad/s\\
$X_{0,1},X_{0,2}$ & $0.1618$ p.u. & $H$ & $1$ s\\
$X_{12}$ & $0.4$ p.u. & $D_p$ & $10$\\
$P_0$ & $0.8$ p.u. & $I_{\max,1},I_{\max,2}$ & $1.2$ p.u.\\
$I_{\mathrm{th}}$ & $1.0$ p.u. & $k_{\mathrm{ins}}$ & $2.35767$\\
\specialrule{0.4pt}{2.5pt}{1.2pt}
\specialrule{0.4pt}{0pt}{0pt}
\end{tabular*}
\end{table}

\subsection{ROA Estimation and Interpretation}
\label{subsec:roa_estimation}
\begin{figure}[!t]
\centering
\includegraphics[width=\columnwidth]{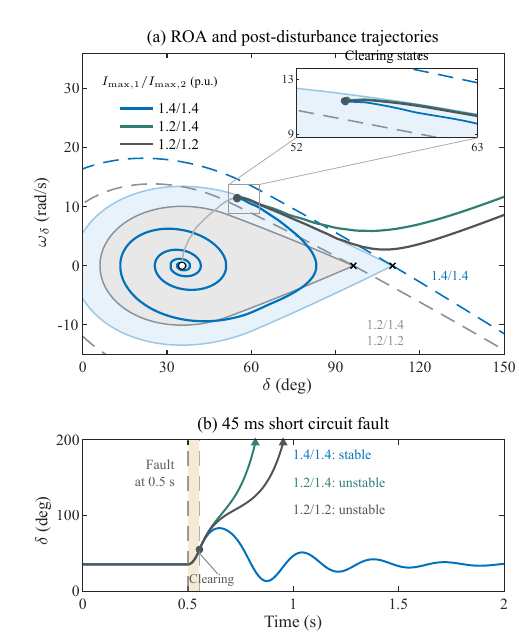}
\caption{Current-limit comparison for the circular current limiter under
a $45$-ms short circuit fault.
(a) ROA boundaries (dashed), energy estimates (shaded), and EMT
trajectories; circles mark clearing states. Blue and gray regions
correspond to minimum current limits of $1.4$ and $1.2$ p.u., respectively.
(b) Relative-angle responses; stable and unstable denote a return to the
target SEP and loss of synchronism, respectively.}
\label{fig:emt_unequal_limits}
\end{figure}

The shaded regions in
Figs.~\ref{fig:roa_energy_certificate},
\ref{fig:emt_time_domain}, and~\ref{fig:emt_unequal_limits}
are the connected components containing the target SEP
of the energy sublevel sets in \eqref{eq:two_gfm_energy_criterion}.
For comparison, numerical ROA boundaries of the reduced angle model
are reconstructed using the stable-manifold method described in
Section~\ref{subsec:physical_insights}, under the applicability
conditions of~\cite{chiang1988stability}. These regions provide the
reference for interpreting the EMT post-jump and fault-clearing states.

The guarantee of convergence to the target SEP applies to the reduced angle model under the
conditions in Section~\ref{subsec:multi_gfm_energy}; the EMT simulations
provide numerical support for its predictions with the implemented
converter controls.

\subsection{Transient Stability Under Phase Jumps and Short-Circuit Faults}
\label{subsec:emt_fault_recovery}
Cases A1--A3 apply retained control-angle offsets $\pm\Delta\delta/2$,
with $\Delta\delta=40^\circ$, $55^\circ$, and $70^\circ$, respectively.
Cases B1 and B2 apply a temporary three-phase-to-ground short-circuit
fault at the GFM1 line-side bus, with clearing commands issued after
$20$ and $40$ ms, respectively.
Fault clearing removes the temporary shunt fault and
restores the original network configuration. The power set points remain unchanged. 
The same post-disturbance ROA is therefore
used to assess the post-jump and fault-clearing states for each case.

These cases compare the transient stability predictions of the energy criterion
in \eqref{eq:two_gfm_energy_criterion} with the EMT responses.
The gray regions in Fig.~\ref{fig:emt_time_domain}(a) and (b) are the energy
sublevel components defined by \eqref{eq:two_gfm_energy_criterion}.
The critical energies calculated from \eqref{eq:two_gfm_critical_energy}
are $\mathcal E_{\mathrm{cr}}=0.16094$ for the circular current limiter
and $0.07583$ for impedance insertion. A1 and B1 lie inside the respective estimates for both
limiters, and A2 also lies inside for the circular current limiter.
The energy criterion guarantees convergence to the target SEP for these
five states in the reduced angle model, and all five EMT trajectories
return to the target SEP, consistent with the ROA estimates obtained
from the energy function in \eqref{eq:relative_energy_dissipation}.

The B2 clearing state with the circular current limiter illustrates the
conservatism of this criterion.
Its energy ratio is $\mathcal E_{\delta}/\mathcal E_{\mathrm{cr}}=1.203$,
placing it outside the energy estimate but inside the reconstructed ROA.
Its subsequent return to the target SEP in EMT is consistent with the additional
dissipation accounted for by the ROA boundary. The energy estimate thus
provides a sufficient transient stability criterion, while the reconstructed ROA
identifies additional initial states from which trajectories converge to the target SEP.

The limiter comparison tests how the critical energy in
\eqref{eq:two_gfm_critical_energy} shapes the transient stability margin.
For the characteristics considered here, impedance insertion lowers
the critical energy and reduces the corresponding energy estimate.
For the $55^\circ$ jump, nearly
identical initial states, with $\delta=90.40^\circ$ and
$\omega_{\delta}\approx0$, lie inside the energy estimate for the
circular current limiter but outside that for impedance insertion.
The latter state also lies outside the reconstructed ROA. The circular current
limiter case returns to the target SEP, whereas impedance insertion leads to
loss of synchronism. For the $40$-ms short circuit fault, both clearing angles are near
$50^\circ$, and the circular current limiter case again returns to the target SEP while
the impedance insertion case loses synchronism. Neither clearing state is certified
by the energy criterion, as both exceed their respective critical energies.
The insets
in Fig.~\ref{fig:emt_time_domain}(a) and (b) locate A2 and B2 relative to the energy estimates and
ROA boundaries.

\subsection{Energy Balance and Transient Stability Margins}
\label{subsec:emt_margin_comparison}
Fig.~\ref{fig:emt_time_domain}(h) tests the dissipation structure
underlying the energy criterion. It compares the decrease in
$\mathcal E_{\delta}$ with the accumulated damping integral,
$\int_{0^+}^{t}D_{\delta}\omega_{\delta}^{2}\,\mathrm d\tau$,
along the stable $40^\circ$ jump trajectories, excluding the imposed jump.
The maximum residuals over the $8$-s records are $3.50\%$ for the circular
current limiter and $2.04\%$ for impedance insertion, normalized by their
post-jump energies. These results
show close agreement with the energy dissipation relation in
\eqref{eq:relative_energy_dissipation}.

Table~\ref{tab:emt_recovery_margins} compares the energy-based critical phase-jump
magnitudes with the EMT transitions between convergence to the target SEP and
loss of synchronism. In the table, ``Circular'' and ``Impedance'' denote the reactive circular
current limiter and impedance insertion, respectively. The thresholds obtained from
\eqref{eq:two_gfm_energy_criterion} agree with the EMT transition
intervals to within $0.85\%$ for the circular current limiter and $0.55\%$
for impedance insertion.
The fault cases certified by the energy criterion also return to the target SEP in EMT.

\begin{table}[!t]
\caption{Energy-based transient stability assessment and EMT results}
\label{tab:emt_recovery_margins}
\centering\footnotesize
\setlength{\tabcolsep}{4pt}
\begin{tabular*}{\columnwidth}{@{\extracolsep{\fill}}lcc@{}}
\specialrule{0.4pt}{0pt}{1.2pt}
\specialrule{0.4pt}{0pt}{2.5pt}
Limiter & \shortstack{Energy-based critical\\phase-jump magnitude (deg)}
 & \shortstack{EMT transition\\interval (deg)}\\
\midrule
Circular & 60.98 & [60.47, 60.55]\\
Impedance & 52.53 & [52.73, 52.81]\\
\specialrule{0.4pt}{2.5pt}{1.2pt}
\specialrule{0.4pt}{0pt}{0pt}
\end{tabular*}
\par\smallskip
\setlength{\tabcolsep}{2.5pt}
\begin{tabular*}{\columnwidth}{@{\extracolsep{\fill}}lrrrrc@{}}
\specialrule{0.4pt}{0pt}{1.2pt}
\specialrule{0.4pt}{0pt}{2.5pt}
Limiter & $T_{\rm cmd}$ & $T_{\rm cl}$ & $\delta_{\rm cl}$ & $\omega_{\delta,\mathrm{cl}}$ & Energy/EMT\\
 & (ms) & (ms) & (deg) & (rad/s) & \\
\midrule
Circular & 20 & 26.5 & 40.23 & 5.71 & Stable/Stable\\
Impedance & 20 & 26.2 & 40.10 & 5.29 & Stable/Stable\\
Circular & 40 & 46.8 & 50.07 & 9.99 & --/Stable\\
Impedance & 40 & 46.4 & 49.77 & 9.49 & --/Unstable\\
\specialrule{0.4pt}{2.5pt}{1.2pt}
\specialrule{0.4pt}{0pt}{0pt}
\end{tabular*}
\par\smallskip
{\footnotesize Stable and Unstable denote a return to the target SEP and loss of synchronism, respectively.
In the energy assessment, -- denotes no conclusion.
Jump brackets run from tested stable to tested unstable cases.}
\end{table}

\subsection{Effect of Unequal Current Limits}
\label{subsec:emt_unequal_limits}

This subsection tests the series composition of two GFMs with unequal current limits.
The results in Section~\ref{subsec:physical_insights} show that,
with unequal current limits for the
circular current limiter,
the stability regions are governed by the lower limit.
Fig.~\ref{fig:emt_unequal_limits} compares current-limit pairs
$(1.4,1.4)$, $(1.2,1.4)$, and $(1.2,1.2)$ p.u. under the same short-circuit
disturbance, with all other settings unchanged.
The latter two configurations share the same theoretical potential,
energy estimate, and ROA because their series current is constrained by
the same lower limit. Both EMT responses exhibit loss of synchronism.
Increasing only one converter's current limit therefore does not remove
the transfer constraint imposed by the other converter.

When both limits are raised to $1.4$ p.u., the electrical potential barrier
increases and the clearing state lies inside the corresponding energy
estimate. Its energy is approximately $0.927\mathcal E_{\mathrm{cr}}$,
so the energy criterion guarantees convergence to the target SEP in the reduced model;
the EMT trajectory also returns to the target SEP.
The other two clearing states lie outside their energy estimate.
These results demonstrate both
the direct use of the constructed energy function for transient stability assessment
and the role of the lower port current limit in determining the transient
stability margin.

\section{Conclusion}
\label{sec:conclusion}

This paper develops a network energy construction that incorporates GFM
current limiting while retaining a provable dissipation property.
For the inductive port and network model considered, the electrical
active powers form the angle gradient of the effective network potential.
Current limiting reshapes this potential, while synchronization damping
dissipates transient energy. The same energy function therefore supports
analysis throughout normal operation, limiter activation, current-limited
operation, and deactivation, without switching between mode-dependent
energy functions. This structure enables energy-based transient stability
evaluation consistent with the classical direct method.

Applying the framework to a two-GFM case provides physical insights
into how current limiting shapes power transfer and transient stability.
It shows that both GFMs can enter current limiting during a transient
and still return to the SEP. With reactive circular current limiting,
the GFM with the lower current capacity sets the bottleneck for transient
stability. Comparisons between the reactive circular
current limiter and impedance insertion reveal how current utilization
shapes the potential barrier and stability margin. EMT simulations
support the predicted stability behaviour and energy dissipation.

Although illustrated here with a two-GFM system, the energy construction
is formulated for networks with an arbitrary number of GFMs under the
stated modelling assumptions. This network formulation provides a
theoretical foundation for scaling energy-based transient stability
assessment to large-scale GFM networks under current limiting.

\bibliographystyle{IEEEtran}
\bibliography{references}

\begin{thebibliography}{10}
\bibitem{zhou_emt_model}
W.~Zhou and Y.~Gu, ``{MATLAB/Simulink} {EMT} model for transient stability of
  grid-forming converter networks with current limiting,'' to be released.
\end{thebibliography}

\end{document}